\documentclass[reqno,12pt]{article}
\usepackage{amsmath,amsfonts,amssymb,amsthm,amstext,amscd,eucal,xcolor}
\usepackage[all]{xy}
\usepackage{amssymb,amsmath,amsfonts,bm,braket,enumerate}
\usepackage{amsmath, amssymb}
\usepackage{mathtools}
\newcommand{\mathsym}[1]{{}} 
\usepackage[colorlinks=true,
            linkcolor=blue,
            urlcolor=blue,
            citecolor=blue]{hyperref}
\usepackage{enumerate}

\usepackage{graphicx}% Include figure files
\usepackage{amsmath} \usepackage{amssymb}
\usepackage{bm}% bold math
\usepackage{cite,hyperref}
\usepackage{mathtools}
\usepackage{amsmath} \usepackage{amssymb}
\usepackage{bm}% bold math
\usepackage{graphicx}
\usepackage[active]{srcltx}
\makeatletter \@addtoreset{equation}{section}

\makeatletter\renewcommand\section{\@startsection {section}{1}{\z@}%
                                   {-3.5ex \@plus -1ex \@minus -.2ex}%nn
                                   {2.3ex \@plus.2ex}%
                                   {\normalfont\large\bfseries}}
\renewcommand\subsection{\@startsection{subsection}{2}{\z@}%
                                     {-3.25ex\@plus -1ex \@minus -.2ex}%
                                     {1.5ex \@plus .2ex}%
                                     {\normalfont\bfseries}}

\newcommand{\be}{\begin{equation}}
\newcommand{\ee}{\end{equation}}
\newcommand{\bea}{\begin{eqnarray}}
\newcommand{\eea}{\end{eqnarray}}
\newcommand{\bse}{\begin{subequations}}
\newcommand{\ese}{\end{subequations}}
\newcommand{\beqa}{\begin{eqnarray}}
\newcommand{\eeqa}{\end{eqnarray}}
\newcommand{\beqar}{\begin{eqnarray*}}
\newcommand{\eeqar}{\end{eqnarray*}}
\newcommand{\bi}{\begin{itemize}}
\newcommand{\ei}{\end{itemize}}
\newcommand{\bn}{\begin{enumerate}}
\newcommand{\en}{\end{enumerate}}
\newcommand{\ba}{\begin{array}}
\newcommand{\ea}{\end{array}}
\newcommand{\bc}{\begin{center}}
\newcommand{\ec}{\end{center}}

\definecolor{darkgreen}{rgb}{0,0.3,0}
\definecolor{darkblue}{rgb}{0,0,0.3}
\definecolor{darkred}{rgb}{0.7,0,0}

\newcommand{\NC}{noncommutative }

\makeatother
\begin{document}

\newcommand{\email}[1]{\footnote{E-mail: \href{mailto:#1}{#1}}}

\title{\bf\large{
Noncommutative Jackiw-Pi model: One-loop renormalization}}

\author{\bf\small{R.~Bufalo} \email{rodrigo.bufalo@dfi.ufla.br} $^{a}$, {M.~Ghasemkhani}\email{ghasemkhani@ipm.ir } $^{b}$ and  M.~Alipour\email{moj.alipour@mail.sbu.ac.ir} $^{b}$ \\
\\
\textit{ \small $^{a}$   Departamento de F\'isica, Universidade Federal de Lavras,}\\
\textit{\small Caixa Postal 3037, 37200-000 Lavras, MG, Brazil}\\
\textit{\small $^{b}$  Department of Physics, Shahid Beheshti University,  G.C., Evin, Tehran 19839, Iran}\\
}
%%%%%%%%%%%%%%%%%%%%%%%%%%%%%%%%%%%%%%%%%%%%%%%%%%%%%%%%%%%%%%%%%%%%%%%%%%%%%%%%%%%%%%%%
%\date{\today}
\maketitle

\begin{abstract}
In this paper, we study the quantum behavior of the noncommutative Jackiw-Pi model. After establishing the Becchi-Rouet-Store-Tyutin (BRST) invariant action, the perturbative renormalizability is discussed, allowing us to introduce the renormalized mass and gauge coupling.
We then proceed to compute the one-loop correction to the basic 1PI functions, necessary to determine the renormalized parameters (mass and charge), next we discuss the physical behavior of these parameters.
\end{abstract}

%\vspace{0.5in}

%%%%%%%%%%%%%%%%%%%%%%%%%%%%%%%%%%%%%%%%%%%%%%%%%%%%%%%%%%%%%%%%%%%%%%%%%%%%%%%%%%%%%%%%%%
\setcounter{footnote}{0}
\renewcommand{\baselinestretch}{1.05}  %Line spacing
%%%%%%%%%%%%%%%%%%%%%%%%%%%%%%%%%%%%%%%%%%%%%%%%%%%%%%%%%%%%%%%%%%%%%%%%%%%%%%%%

%\addtocontents{toc}{\protect\setcounter{tocdepth}{2}}
\newpage
\tableofcontents
%%%%%%%%%%%%%%%%%%%%%%%%%%%%%%%%%%%%%%%%%%%%%%%%%%%%%%%%%%%%%%%%%%%%%%%%%%%%%%%%%%%%%%%%%%%%%%%%%%%%%%%%%%%%%%%%%%%%%%%%%%%%%%%%%%%%%%%%%%%
\newpage
\section{Introduction}
\label{sec1}

Mass generation for quantum fields has always been an important subject extensively studied even after the establishment of standard model by means of Higgs mechanism.
Interestingly on its own, mass generation is seen by a completely new optics when the dimensionality of space-time is lowered to $(1+1)$ and $(2+1)$ dimensions. Mainly, in such cases there is a compatibility between gauge symmetry and massive vector fields, where nonperturbative effects play an important role and topological terms are allowed, respectively.
The first example of the presence of a massive photon
without breaking the gauge symmetry is the toy model QED$_2$, the so-called Schwinger model \cite{schwinger1,schwinger2}.
Furthermore, gauge field theories when defined in three space-time dimensions carry notorious attention since the early
works of Deser, Jackiw and Templeton \cite{jackiw}.
These $(2+1)$-dimensional field  models possess not only interesting mathematical structure in their solution, but rather they are well motivated by allowing a gauge field theoretical description of (planar) condensed matter phenomena, such as high-$T_c$ superconductivity and quantum Hall effect, among other examples \cite{Marino:2017ckg}.

In more details, the Chern-Simons term is a topological theory which when is added to the three-dimensional Maxwell/Yang-Mills action, renders the gauge field a massive mode, while preserving gauge invariance.
However, the price to pay due to the presence of a Chern-Simons topological mass term is the violation of parity-invariance.
On the other hand, if parity invariance is required to be preserved, one might approach this mass-gap generation mechanism through the doublet mechanism.
Using this method Jackiw and Pi have suggested a theory for massive
vector fields, which is simultaneously gauge invariant and parity preserving, this is namely the Jackiw-Pi model \cite{Jackiw:1997jga,Jackiw:1997ha}.
In this case, the two vector fields have opposite parity transformations, which generate a mass-gap through a mixed Chern-Simons-like term preserving parity; moreover, the parity transformation is defined to include a field exchange together with the coordinate reflection, and this is a symmetry of the doubled theory.
Many aspects concerning the BRST quantization of the Jackiw-Pi model have been studied \cite{Henneaux:1997mf,Dayi:1997in,DelCima:2011bx,Gupta:2011cta,DelCima:2012bm,Kumar:2015mnp, Nishino:2015hha,Nikoofard:2016tzz}.
Actually, a subtle point concerning the Jackiw-Pi model is that it possess two independent local gauge symmetries (inherent due to the doublet mechanism), this clash among symmetries is known as bifurcation effect \cite{Deser:2012ci}.
We shall revisit this point from a BRST point of view as presented in Ref.~\cite{DelCima:2011bx}.

Recently a proposal of extending Jackiw-Pi model to a noncommutative spacetime has been presented \cite{Nikoofard:2016tzz}.
There the Batalin-Vilkovisky formalism has been used in conjunction with the enveloping algebra approach for non-Abelian noncommutative field theory \cite{Jurco:2001rq} to give a proper formulation for quantization.
Because of the recent improvement concerning the precision
of the measurements of experiments (LHC, ILC, etc.)
investigating particle properties in the search of direct evidence of new physics, noncommutative (NC) gauge field theory has received significant attention due to its interesting way to engender Lorentz violating effects and also by its richer phenomenological aspects \cite{ref16,ref17,ref18,ref23}.
One may also say that noncommutativity of the coordinates of the spacetime emerges naturally in the description of fractional quantum Hall effect \cite{fradkin}, and find application in the study of planar
physics in condensed matter and statistical physics.

It is well known that the Maxwell-Chern-Simons theory, like the Jackiw-Pi model, is UV finite, while its NC version exhibits UV/IR mixing at one-loop order \cite{Ghasemkhani:2015tqu}.
The UV/IR effect is one of the main drawback features of the NC field theories,
so due to the potential application of Jackiw-Pi model into planar physical systems with parity symmetry, it is rather interesting to explore theoretical aspects and studying whether we still encounter this IR instabilities in the NC Jackiw-Pi model. Also, an underlying question in the analysis is that whether
a parity even NC field theory is sufficient to render a NC gauge model free of noncommutative IR instabilities, i.e. an additional discrete symmetry can change significantly the physical behavior of a field theory.
We wish to present a detailed account for the BRST renormalizability of the model from the point of view of Ref.~\cite{DelCima:2011bx}, and also to compute the first order perturbative correction for the basic 1PI functions.
This paper is organized as follows.
In Sec.~\ref{sec2} we present an overview on the Jackiw-Pi model, exploring the presence of a double Abelian symmetry, or bifurcation effect, and the resulting BRST structure used for the gauge fixing and ghosts. We extend this BRST description to the noncommutative case, where the non-Abelian structure is replaced by the Moyal star product\footnote{The noncommutativity we will be using in the paper is defined by the algebra $[\hat{x}_\mu , \hat{x}_\nu]=i\theta_{\mu \nu}$. So to construct a noncommutative field theory, using the Weyl-Moyal (symbol) correspondence, allowing to the ordinary product be replaced by the Moyal star product as defined below.}. Discrete symmetries in the NC Jackiw-Pi model is also discussed.
In Sec.~\ref{sec3} all the Feynman rules are presented for the propagators and 1PI vertices, in addition, a discussion for the renormalizability for the full theory is presented.
Section \ref{sec4} is devoted to present and compute the graphs corresponding to the one-loop self-energy functions necessary to determine the renormalized mass and coupling constant.
In Sec.~\ref{sec5} the expressions for the finite counter-terms are computed, as well as we discuss the physical behavior of the renormalized mass and coupling constant.
Final remarks are presented in Sec.~\ref{conc}.

%%%%%%%%%%%%%%%%%%%%%%%%%%%%%%%%%%%%%%%%%%%%%%%%%%%%%%%%%%%%%%%%%%%%%%%%%
%%%%%%%%%%%%%%%%%%%%%%%%%%%%%%%%%%%%%%%%%%%%%%%%%%%%%%%%%%%%%%%%%%%%%%%%%%%%%%%%%%%%%%%%%%%%%%%%

\section{Jackiw-Pi model}
\label{sec2}

Before starting with the noncommutative extension of the Jackiw-Pi model, let us briefly review the main parts of its construction, so that it would help us to highlight some important points in the noncommutative construction.
The Jackiw-Pi model is a non-Abelian gauge invariant, mass generating,
parity preserving theory whose dynamics is governed by the Lagrangian
\begin{equation}
\mathcal{L}=Tr\Big(-\frac{1}{2}F_{\mu\nu}F^{\mu\nu}-\frac{1}{2}G_{\mu\nu}G^{\mu\nu}
+m\epsilon^{\mu\nu\lambda}F_{\mu\nu}\phi_{\lambda}\Big),\label{eq:0.1}
\end{equation}
where $A_{\mu}$ and $\phi_{\mu}$ are parity even and odd vector
bosonic fields, respectively, and $m$ is a mass parameter.
This Lagrangian can be understood as describing a charged vector mesons $\phi_{\mu}$ minimally coupled with a gauge potential $A_{\mu}$, where the fields have opposite parity transformations, which generates a mass-gap through a mixed Chern-Simons-like term preserving parity.
Moreover, the two-form curvatures and covariant derivative are given as
\begin{align}
F_{\mu\nu}& =\partial_{\mu}A_{\nu}-\partial_{\nu}A_{\mu}+ig\left[A_{\mu},A_{\nu}\right],\\
G_{\mu\nu}& =D_{\mu}\phi_{\nu}-D_{\nu}\phi_{\mu},
\end{align}
where $D_{\mu}\bullet=\partial_{\mu}\bullet+ig\left[A_{\mu},\bullet\right]$. It is worth mentioning that the full nonlinear theory \eqref{eq:0.1}
does have an interesting symmetry structure. In addition to the (Yang-Mills) gauge symmetry
\begin{equation}
\delta_{\theta}A_{\mu}=D_{\mu}\theta,\quad\delta_{\theta}\phi_{\mu}=ig\left[\phi_{\mu},\theta\right].
\end{equation}
The (massive) mixing term in $\mathcal{L}$ is also invariant
upon the (non-Yang-Mills) symmetry
\begin{equation}
\delta_{\chi}A_{\mu}=0,\quad\delta_{\chi}\phi_{\mu}=D_{\mu}\chi.
\end{equation}
However, this second transformation does not leave the nonlinear part
of $G_{\mu\nu}$ invariant, since $\delta_{\chi}G_{\mu\nu}=\left[F_{\mu\nu},\chi\right]$.
In other words, this shows that the quadratic theory possesses two independent, Abelian gauge symmetries; while with interaction, only one non-Abelian symmetry survives \cite{Jackiw:1997jga,DelCima:2012bm}.
This clash among the two local Abelian invariance symmetries is the so-called bifurcation effect \cite{Deser:2012ci}.
This presents an intricate quantization problem that is
solved by enlarging the gauge symmetry with an additional scalar field
$\rho$, that transforms accordingly
\begin{align}
\delta_{\theta}\rho & =ig\left[\rho,\theta\right]\nonumber \\
\delta_{\chi}\rho & =-\chi. \label{eq:0.5}
\end{align}
So that, the replacement
\begin{equation}
G_{\mu\nu}^{a}\rightarrow G_{\mu\nu}^{a}+f_{abc}F_{\mu\nu}^{b}\rho^{c},
\end{equation}
in the Lagrangian density \eqref{eq:0.1} allows Hamiltonian path integral quantization \cite{Jackiw:1997jga,DelCima:2012bm}.
It is even possible that in addition to the above replacement, a kinetic term for the $\rho$ field is present in the form $Tr\left( D_\mu \rho - m \phi_\mu \right)^2$, this new model is the so-called extended Jackiw-Pi model \cite{Nishino:2015hha}.

The corresponding BRST transformations of these fields, stemmed from
these gauge transformations ($\delta_{\theta}$ and $\delta_{\chi}$),
are \cite{DelCima:2012bm}
\begin{align}
sA_{\mu} & =D_{\mu}c,\quad sc=-gc^{2},\quad s\overline{c}=0,\quad sb=0,\quad s\rho=-\xi+ig\left[\rho,c\right] \\
s\phi_{\mu} & =D_{\mu}\xi+ig\left[\phi_{\mu},c\right],\quad s\xi=-g\left[\xi,c\right],\quad s\overline{\xi}=0,\quad s\pi=0,\label{eq:0.3}
\end{align}
where $c$ and $\xi$ are two sorts of Faddeev-Popov ghosts, while $\overline{c}$ and $\overline{\xi}$ Faddeev-Popov antighosts, and $b$ and $\pi$ are Nakanishi-Lautrup auxiliary fields.

The gauge fixing is then obtained in the BRST formalism as usual, and reads
\begin{equation}
\mathcal{L}_{\rm g.f}=Tr\Big(b\partial_{\mu}A^{\mu}-\overline{c}\partial^{\mu}D_{\mu}c+\pi\partial_{\mu}\phi^{\mu}
-\overline{\xi}\partial^{\mu}\left(D_{\mu}\xi+ig\left[\phi_{\mu},c\right]\right)+\frac{\alpha}{2}b^{2}
+\frac{\beta}{2}\pi^{2}\Big).\label{eq:0.4}
\end{equation}
Additionally, due to the transformations \eqref{eq:0.5} we can also consider the gauge fixing $\rho=0$, without loss of generality.
In this sense, using this BRST description, all the vector fields have nonsingular behavior, and their propagators can suitably be defined \cite{DelCima:2012bm}.

\subsection{Noncommutative framework}

The noncommutative extension of the Jackiw-Pi model is defined by the
following Lagrangian density \cite{Nikoofard:2016tzz}
\begin{equation}
\mathcal{L}=-\frac{1}{4}F_{\mu\nu}\star F^{\mu\nu}-\frac{1}{4}\left(G_{\mu\nu}+ig\left[F_{\mu\nu},\rho\right]_{\star}\right)\star\left(G^{\mu\nu}
+ig\left[F^{\mu\nu},\rho\right]_{\star}\right)+\frac{m}{2}\epsilon^{\mu\nu\lambda}F_{\mu\nu}\star\phi_{\lambda},\label{eq:1.0}
\end{equation}
where $\left[~,~\right]_{\star}$ is the Moyal bracket. Notice that this Lagrangian can be understood as before describing a charged vector mesons $\phi_{\mu}$ minimally coupled with a gauge potential $A_{\mu}$, being the main difference between the two forms due to the presence of the Moyal star product engendering the non-Abelian interaction structure. The Moyal star product between functions is described as
\begin{equation}
f\left( x\right) \star g\left(x\right) = f\left( x\right) \exp \Big(\frac{i}{2}\theta ^{\mu \nu}
\overleftarrow{\partial_\mu}
 \overrightarrow{\partial_\nu}\Big) g\left( x\right).
\end{equation}
The two-form curvatures in this setup are given as
\begin{align}
F_{\mu\nu} & =\partial_{\mu}A_{\nu}-\partial_{\nu}A_{\mu}+ig\left[A_{\mu},A_{\nu}\right]_{\star},\\
G_{\mu\nu} & =\nabla_{\mu}\phi_{\nu}-\nabla_{\nu}\phi_{\mu},
\end{align}
where $\nabla_{\mu}\bullet =\partial_{\mu}\bullet+ig\left[A_{\mu},\bullet\right]_{\star}$. As we have discussed, we can extend the original structure \eqref{eq:0.3} and establish the corresponding BRST transformations of this noncommutative setup,
\begin{align}
\mathfrak{s} A_{\mu} & =\nabla_{\mu}c,\quad \mathfrak{s}c=-gc^{2},\quad \mathfrak{s}\overline{c}=0,\quad \mathfrak{s}b=0,\quad \mathfrak{s}\rho=-\xi+ig\left[\rho,c\right]_{\star}\\
\mathfrak{s}\phi_{\mu} & =\nabla_{\mu}\xi+ig\left[\phi_{\mu},c\right]_{\star},\quad \mathfrak{s}\xi=-g\left[\xi,c\right]_{\star},\quad \mathfrak{s}\overline{\xi}=0,\quad \mathfrak{s}\pi=0,
\end{align}
where the ghosts, antighosts and auxiliary fields are the same as before in the non-Abelian setup Eq.~\eqref{eq:0.3}.
Finally, the gauge fixing is obtained in the BRST formalism, and reads
\begin{align}
\mathcal{L}_{\rm g.f}&= \mathfrak{s}\Big( \overline{c} \partial^{\mu} A_\mu +\overline{\xi} \partial^\mu \phi_\mu +\frac{\alpha}{2}\overline{c} b
+\frac{\beta}{2}\overline{\xi} \pi \Big)\nonumber \\
&=b\star\partial_{\mu}A^{\mu}+\partial^{\mu}\overline{c}\star \nabla_{\mu}c+\pi\star\partial_{\mu}\phi^{\mu}+\partial^{\mu}\overline{\xi}\star\left(\nabla_{\mu}\xi+ig
\left[\phi_{\mu},c\right]_{\star}\right)+\frac{\alpha}{2}b\star b+\frac{\beta}{2}\pi\star\pi. \label{eq:1.1}
\end{align}
Once again, we can consider the gauge fixing $\rho=0$, without
loss of generality.
This construction assures that all the vector fields have a well-defined structure, so that we can proceed to the computation of propagators and vertex functions.

\subsubsection{Discrete symmetries}

In order to have a full view of the NC Jackiw-Pi model, we
shall next analyze the behavior of the Lagrangian density
\eqref{eq:1.0} under discrete symmetries: parity, charge conjugation and time reversal. This is also motivated because Jackiw-Pi model is seen as a parity invariant extension of Chern-Simons theory, so it is crucial to establish under which conditions this holds on a noncommutative spacetime.
It should be emphasized that the algebra $[x^{\mu},x^{\nu}] = i\theta^{\mu\nu} $, with the assumption that $\theta^{0i}=0$, plays an important role in the present analysis.

\begin{itemize}

  \item [(\emph{i})]~\emph{Parity}

Parity transformation in $2+1$ dimensions is
indeed a reflection described by $x_1 \rightarrow - x_1$ and
$x_2 \rightarrow x_2$. Under parity, the gauge field transforms as
\begin{equation}
A_0 \rightarrow A_0, \quad A_1 \rightarrow -A_1, \quad A_2 \rightarrow A_2,
\end{equation}
Now under the change $\theta_{ij} \rightarrow - \theta_{ij}$ we find that $F_{\mu \nu}F^{\mu \nu}$ is parity invariant.
Furthermore, with this additional condition, and by imposing that the vector field $\phi_\mu$ transforms as
\begin{equation}
\phi_0 \rightarrow -\phi_0, \quad \phi_1 \rightarrow \phi_1, \quad \phi_2 \rightarrow -\phi_2,
\end{equation}
it is easy to establish that the remaining terms of \eqref{eq:1.0} are parity invariant also in the noncommutative case.

  \item  [(\emph{ii})]~\emph{Charge conjugation}

  Under a charge conjugation transformation, the
gauge field changes as $A_\mu \rightarrow - A_\mu$. However, one can realize that the noncommutative Maxwell term is not C-invariant unless we consider $\theta_{ij} \rightarrow - \theta_{ij}$.
Now the remaining interacting terms between $A_\mu$ and $\phi_\mu$ in the Lagrangian density \eqref{eq:1.0} are invariant under charge conjugation if the charged vector field transforms as $\phi_\mu \rightarrow - \phi_\mu$. Establishing thus the conditions that leave the full Lagrangian C-invariant.

    \item [(\emph{iii})]~\emph{Time reversal}

    Under a time reversal transformation, $x_0 \rightarrow -x_0$, the gauge
field now changes as
\begin{equation}
A_0 \rightarrow A_0, \quad A_1 \rightarrow -A_1, \quad A_2 \rightarrow -A_2,
\end{equation}
and with the change $\theta_{ij} \rightarrow - \theta_{ij}$, one see that the NC Maxwell's term is T-invariant.
Additionally, imposing that the vector field $\phi_\mu$ transforms under the time reversal as
\begin{equation}
\phi_0 \rightarrow -\phi_0, \quad \phi_1 \rightarrow \phi_1, \quad \phi_2 \rightarrow \phi_2,
\end{equation}
the remaining interacting terms between $A_\mu$ and $\phi_\mu$ in the Lagrangian density \eqref{eq:1.0} are shown to be invariant under time reversal.

\end{itemize}

This analysis shows that the NC Jackiw-Pi model is invariant under all discrete symmetries, this result is in contrast with the usual NC Maxwell-Chern-Simons model \cite{Ghasemkhani:2015tqu}, enlarging thus the possibilities of applications. The behavior of the NC Jackiw-Pi model under discrete symmetries can be summarized as the following:
\begin{center}
\begin{tabular}{|c|c|c|c|c|c|c|}
\hline
term & P & C & T & CP & PT & CPT \\
\hline
$F_{\mu\nu}F^{\mu\nu}$ & + & + & + & + & + & + \\
\hline
$G_{\mu\nu}G^{\mu\nu}$ & + & + & + & + & + & + \\
\hline
$\epsilon^{\mu\nu\rho}F_{\mu\nu}\phi_{\rho}$& + & + & + & + & + & + \\ \hline
\end{tabular}
\end{center}
%%%%%%%%%%%%%%%%%%%%%%%%%%%%%%%%%%%%%%%%%%%%%%%
%%%%%%%%%%%%%%%%%%%%%%%%%%%%%%%%%%%%%%%%%%%%%%%

\section{Propagators and Renormalizability}
\label{sec3}

As it is well known, from the functional methods, the 1PI function is
given in terms of the connected function as
\[
\Gamma\left[\Phi_{i}\right]=W\left[J_{i}\right]-\int d^{3}x~J_{i}(x)\Phi_{i}(x),
\]
where $\Phi_{i}$ denotes collectively the whole set of fields, and $J_{i}$ the respective set of currents.
Moreover, at zeroth order, the effective action is precisely the free action $\Gamma^{\left(0\right)}\left[\Phi_{i}\right]=\int d^{3}x\left(\mathcal{L}+\mathcal{L}_{\rm g.f}\right)$.
From these relations we find that the tree-level propagators at momentum space read
\begin{align}
iD_{\mu\nu}\left(k\right) & =\frac{1}{k^{2}-m^{2}}\Big(\eta_{\mu\nu}-\frac{k_{\mu}k_{\nu}}{k^{2}}\Big)-\frac{\alpha}{k^{2}}
\frac{k_{\mu}k_{\nu}}{k^{2}},\label{eq:0.13}\\
iS_{\mu\nu}\left(k\right) & =\frac{1}{k^{2}-m^{2}}\Big(\eta_{\mu\nu}-\frac{k_{\mu}k_{\nu}}{k^{2}}\Big)
-\frac{\beta}{k^{2}}\frac{k_{\mu}k_{\nu}}{k^{2}},\label{eq:0.14}\\
iW_{\mu\nu}\left(k\right) & =-i\frac{m}{k^{2}\left(k^{2}-m^{2}\right)}\epsilon_{\mu\nu\lambda}k^{\lambda},\label{eq:0.15}
\end{align}
where the propagators $D_{\mu\nu}$, $S_{\mu\nu}$ and $W_{\mu\nu}$ are related with the vev's $\left\langle A_{\mu}A_{\nu}\right\rangle $,
$\left\langle \phi_{\mu}\phi_{\nu}\right\rangle $ and $\left\langle A^{\mu}\phi^{\nu}\right\rangle $, respectively.
Moreover, for the ghost fields $\left\langle \overline{c} c\right\rangle $ and $\left\langle \overline{\xi} \xi\right\rangle$ are given by
\begin{equation}
D_{c}\left(p\right)=D_{\xi}\left(p\right)=\frac{i}{p^{2}}.\label{eq:0.16}
\end{equation}
The respective vertex Feynman rules are below \footnote{These $n$-point vertex functions have an implicit energy-momentum conservation constraint $\delta^{(3)}\left(p_{1}+\cdots+p_{n}\right)$. }
\begin{itemize}
\item The three gauge field vertex $\left\langle A_{\mu}A_{\nu}A_{\sigma}\right\rangle $
%\end{itemize}
\begin{align}
\Psi^{\mu\nu\sigma}\left(p_{1},p_{2},p_{3}\right) & =2ig \Big[\eta^{\sigma\mu}\left(p_{1}-p_{3}\right)^{\nu}+\eta^{\nu\sigma}\left(p_{3}-p_{2}\right)^{\mu}+
\eta^{\mu\nu}\left(p_{2}-p_{1}\right)^{\sigma}\Big]\sin\Big(\frac{1}{2}\left(p_{1}\times p_{2}\right)\Big)
\end{align}

%\begin{itemize}
\item The four gauge field vertex $\left\langle A_{\mu}A_{\nu}A_{\sigma}A_{\rho}\right\rangle $
\begin{align}
\Psi^{\mu\nu\sigma\rho}\left(p_{1},p_{2},p_{3},p_{4}\right) & =-4g^{2}\Big[\left(\eta^{\mu\sigma}\eta^{\nu\rho}-\eta^{\mu\rho}\eta^{\nu\sigma}\right)\sin\Big(\frac{1}{2}\left(p_{1}\times p_{2}\right)\Big)\sin\Big(\frac{1}{2}\left(p_{3}\times p_{4}\right)\Big)\nonumber \\
 & +\left(\eta^{\nu\sigma}\eta^{\mu\rho}-\eta^{\sigma\rho}\eta^{\nu\mu}\right)\sin\left(\frac{1}{2}\left(p_{3}\times p_{1}\right)\right)\sin\Big(\frac{1}{2}\left(p_{2}\times p_{4}\right)\Big)\nonumber \\
 & +\left(\eta^{\sigma\rho}\eta^{\nu\mu}-\eta^{\nu\rho}\eta^{\mu\sigma}\right)\sin\Big(\frac{1}{2}\left(p_{1}\times p_{4}\right)\Big)\sin\Big(\frac{1}{2}\left(p_{2}\times p_{3}\right)\Big)\Big]
\end{align}

\item The two charged and one gauge field vertex $\left\langle A_{\mu}\phi_{\nu}\phi_{\sigma}\right\rangle $
\begin{align}
\Upsilon^{\mu\nu\sigma}\left(p_{1},p_{2},p_{3}\right)=2ig \Big[\left(p_{2}\right)^{\sigma}\eta^{\mu\nu}+\left(p_{3}-p_{2}\right)^{\mu}\eta^{\sigma\nu}
-\left(p_{3}\right)^{\nu}\eta^{\mu\sigma}\Big]\sin\Big(\frac{1}{2}\left(p_{1}\times p_{2}\right)\Big)
\end{align}

\item The two charged and two gauge fields vertex $\left\langle A_{\mu}A_{\nu}\phi_{\sigma}\phi_{\rho}\right\rangle $
\begin{align}
\Upsilon^{\mu\nu\sigma\rho}\left(p_{1},p_{2},p_{3},p_{4}\right) & =-4g^{2} \Big[\left(\eta^{\mu\nu}\eta^{\sigma\rho}-\eta^{\sigma\nu}\eta^{\mu\rho}\right)\sin\Big(\frac{1}{2}\left(p_{1}\times p_{3}\right)\Big)\sin\Big(\frac{1}{2}\left(p_{2}\times p_{4}\right)\Big) \nonumber \\
 & +\left(\eta^{\mu\nu}\eta^{\rho\sigma}-\eta^{\rho\nu}\eta^{\mu\sigma}\right)\sin\Big(\frac{1}{2}\left(p_{1}\times p_{4}\right)\Big)\sin\Big(\frac{1}{2}\left(p_{2}\times p_{3}\right)\Big)\Big]
\end{align}

\item The one charged and two gauge fields vertex $\left\langle A_{\mu}A_{\nu}\phi_{\sigma}\right\rangle $
\begin{align}
\Gamma^{\mu\nu\sigma}\left(p_{1},p_{2},p_{3}\right)=2mg\epsilon^{\mu\nu\sigma} \sin\Big(\frac{1}{2}\left(p_{1}\times p_{2}\right)\Big)
\end{align}

\item The two ghosts and one gauge fields vertex $\left\langle A_{\mu}\overline{c}c\right\rangle =\left\langle A_{\mu}\overline{\xi}\xi\right\rangle $
\begin{align}
\Psi^{\mu}\left(p_{2},p_{3}\right)=2ig\Big[\left(p_{2}\right)^{\mu}\sin\Big(\frac{1}{2}\left(p_{2}\times p_{3}\right)\Big)\Big].
\end{align}

\item The mixed two ghosts and one charged fields vertex $\left\langle \phi_{\mu}\overline{\xi}c\right\rangle $
\begin{align}
\Delta^{\mu}\left(p_{2},p_{3}\right)=2ig\Big[\left(p_{2}\right)^{\mu}\sin\Big(\frac{1}{2}\left(p_{2}\times p_{3}\right)\Big)\Big].
\end{align}

\end{itemize}

We shall next proceed in establishing the one-loop renormalization of the noncommutative model by making use of the Slavnov-Taylor identities among the Green's functions and also the universality of the gauge coupling.

\subsection{Renormalizability analysis}

Let us focus on the interacting part of the NC Jackiw-Pi model coming from Eqs.~\eqref{eq:1.0} and \eqref{eq:1.1}
\begin{align}
\mathcal{L}_{\text{int}} & =-ig\partial_{\mu}A_{\nu}\star\left[A^{\mu},A^{\nu}\right]_{\star}+\frac{g^{2}}{4}\left[A^{\mu},A^{\nu}\right]_{\star}\star\left[A_{\mu},A_{\nu}\right]_{\star} +\frac{igm}{2}\epsilon^{\mu\nu\lambda}\left[A_{\mu},A_{\nu}\right]_{\star}\star\phi_{\lambda}\nonumber \\
 & -ig \left(\partial_\mu \phi_\nu -\partial_\nu \phi_\mu \right)\star \left[A^{\mu},\phi^{\nu}\right]_{\star}
 + \frac{g^2}{2}\left(\left[A_{\mu},\phi_{\nu}\right]_{\star} - \left[A_{\nu},\phi_{\mu}\right]_{\star} \right)\star \left[A^{\mu},\phi^{\nu}\right]_{\star}  \nonumber \\
 & +ig\partial^{\mu}\bar{c}\star\left[A_{\mu},c\right]_{\star}+ig\partial^{\mu}\bar{\xi}
 \star\left[A_{\mu},\xi\right]_{\star}+ig\partial^{\mu}\bar{\xi}\star\left[\phi_{\mu},c\right]_{\star},
\end{align}
which is BRST invariant by construction.
In order to check the universality of the gauge coupling by renormalization, which is a result of the Slavnov-Taylor identities, we start by introducing the renormalized fields as below
\[
A_{\mu}^{\left(0\right)}=\sqrt{Z_{3}}A_{\mu},\quad\phi_{\mu}^{\left(0\right)}=\sqrt{Z_{2}}\phi_{\mu},\quad c_{\mu}^{\left(0\right)}=\sqrt{\tilde{Z}_{3}}c,\quad\xi^{\left(0\right)}=\sqrt{\tilde{Z}_{2}}\xi,
\].

The first point to note before proceeding is that the Chern-Simons
coupling $m$ is also renormalized from the mixing propagator $\left\langle A\phi\right\rangle $,
which gives
\begin{equation}
m\sqrt{Z_{2} Z_{3}}=m_{\rm ren}Z_{m},\label{eq:1}
\end{equation}
which defines the renormalization constant $Z_{m}$ related to the
parameter $m_{\rm ren}$. Introducing renormalization constants for the basic vertices we have that $Z_{3A}$, $Z_{4A}$, $Z_{1}$, $\tilde{Z}_{1}$, $\tilde{Z}_{4}$, $\tilde{Z}_{3}^{\rm gh}$, $\tilde{Z}_{2}^{\rm gh}$, and $\tilde{Z}_{4}^{\rm gh}$  are related respectively to the $\left\langle AAA\right\rangle $, $\left\langle AAAA\right\rangle$, $\left\langle \phi AA\right\rangle$, $\left\langle \phi \phi A\right\rangle$, $\left\langle \phi \phi AA\right\rangle$, $\left\langle \bar{c}cA\right\rangle$, $\left\langle \bar{\xi}\xi A\right\rangle$, and $\left\langle \bar{\xi}c\phi\right\rangle$ vertex functions.

We note that the universality of the coupling constants is used when
defining them, i.e. different vertex functions couple with the same
gauge coupling. Relations among these constants follow from the Slavnov-Taylor identities that can be casted simply into the form
\begin{equation}
\frac{Z_{4A}}{Z_{3A}}=\frac{Z_{3A}}{Z_{3}}=\frac{Z_{1}}{Z_{m}}=\frac{\tilde{Z}_{3}^{\rm gh}}{\tilde{Z}_{3}}=\frac{\tilde{Z}_{2}^{\rm gh}}{\tilde{Z}_{2}} = \frac{\tilde{Z}_1}{Z_2}= \frac{\tilde{Z}_4}{\tilde{Z}_1}.\label{eq:8}
\end{equation}
Notice though that the constant $\tilde{Z}_{4}^{\rm gh}$ cannot
be exactly related to others like in \eqref{eq:8}, it can be expressed as
\begin{equation}
\tilde{Z}_{4}^{\rm gh}=\sqrt{\frac{Z_{2}\tilde{Z}_{2}^{\rm gh}\tilde{Z}_{3}^{\rm gh}}{Z_{3}}},
\end{equation}
being then determined by means of other basic constants. In conclusion, to compute the coupling constant renormalization
\begin{equation*}
g_{\rm ren}=Z_{g}g.
\end{equation*}
It is convenient to consider the form
\begin{equation}
Z_{g}=\frac{Z_{3}^{1/2}\tilde{Z}_{3}}{\tilde{Z}_{3}^{\rm gh}}.
\end{equation}
This is the simplest form for the coupling renormalization constant
in the absence of the matter fields, which takes into account the least number of graphs for the corrections associated with the propagators $\left\langle AA\right\rangle$ and $\left\langle \bar{c}c\right\rangle$, and $\left\langle \bar{c}cA\right\rangle$ vertex function.
On the other hand, to determine the renormalized mass $m_{\rm ren}=\frac{\sqrt{Z_{2} Z_{3}} }{Z_m} m$ we should additionally compute corrections to the propagators $\left\langle \phi\phi \right\rangle$ and $\left\langle  A\phi\right\rangle$.
Nonetheless, we shall focus on the discussion about the corrections to the tree-level propagators and subsequently onto the renormalization of the gauge coupling $g$ and mass parameter $m$.

%%%%%%%%%%%%%%%%%%%%%%%%%%%%%%%%%%%%%%%%%%%%%%%%%%%%%%%%%%%%%%%%%%%%%%%
%%%%%%%%%%%%%%%%%%%%%%%%%%%%%%%%%%%%%%%%%%%%%%%%%%%%%%%%%%%%%%%%%%%%%%%

\section{Radiative corrections}
 \label{sec4}

Having derived the relevant Feynman rules, we shall now proceed on the computation of the one-loop radiative corrections to the basic 1PI functions.
Due to the highly intricate form of these momentum dependent functions,
which are rather difficult to compute exactly (and no substantial
information would be obtained), we will consider the low-energy limit so that these expressions can be computed analytically.
This is also known as the highly noncommutative limit, i.e., $p^2/m^2 \rightarrow 0$ while $\tilde{p}$ is kept finite. Moreover, we shall consider Landau gauge on our analysis, which amounts to take $\alpha =0$ and $\beta=0$.

 \subsection{One-loop $\left\langle A A\right\rangle$ self-energy}

   \begin{figure}[t]
\vspace{-1.8cm}
\includegraphics{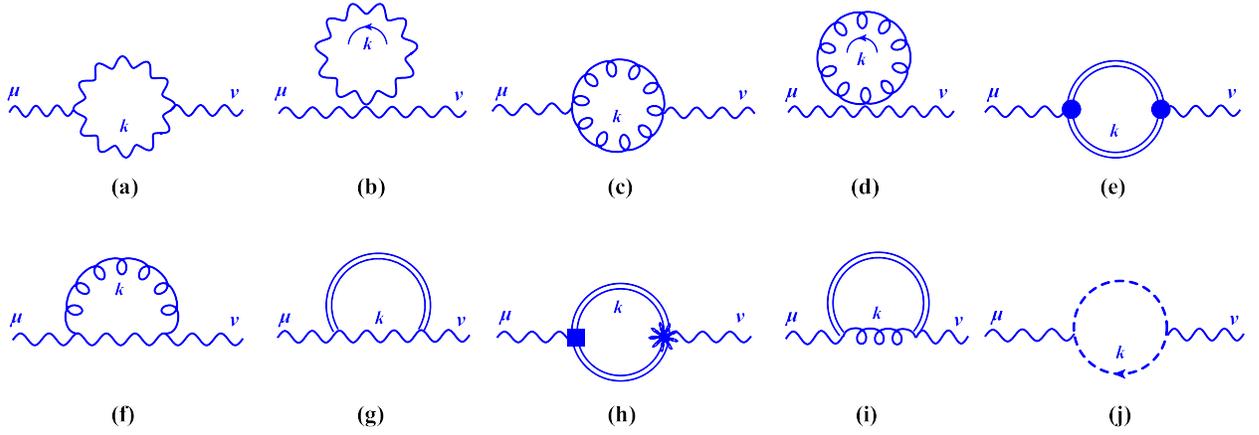}
 \centering\caption{Graphs (a) to (j) represent the one-loop contributions to the self-energy of the gauge field.
 Wavy lines stand for the gauge field $A_\mu$, curly lines for the charged vector field $\phi_\mu$, double solid lines for the mixed propagator $\langle A\phi\rangle$, and dashed lines stand for ghost fields.
In graphs (e) and (h) a circle represents the $\langle AA\phi\rangle$ vertex, a square the $\langle AAA\rangle$ vertex, and a star the $\langle A\phi\phi\rangle$ vertex, such vertex functions can be permuted giving rise to a new contribution.}
\label{fig1}
\end{figure}

We have eleven graphs contributing to the photon polarization tensor at one-loop order; these are depicted in Fig.~\ref{fig1}.
All the contributions can easily be written using the aforementioned Feynman rules, also they have a similar structure that can be cast into a simple form as
 \begin{align}
\Pi^{\mu\nu}(p) = 2g^{2} \int \frac{d^{d}k}{(2\pi)^{d}} \frac{N^{\mu\nu}}{[(p+k)^{2}-m^{2}][k^{2}-m^{2}](p+k)^{2}k^{2}} \sin^{2}\big(\frac{p \times k}{2}\big), \label{4.1}
\end{align}
being the tensor structure on the numerator consists of the eleven contributions
\begin{align}
N^{\mu\nu}=N^{\mu\nu}_{(a)} + N^{\mu\nu}_{(c)} - 2 \Big(N^{\mu\nu}_{(b)} + N^{\mu\nu}_{(d)}\Big)  +m^{4}N^{\mu\nu}_{(e)}
- m^2\Big(N^{\mu\nu}_{(f)} +N^{\mu\nu}_{(g)}+N^{\mu\nu}_{(h,1)}+N^{\mu\nu}_{(h,2)}+N^{\mu\nu}_{(i)}\Big)- 4N^{\mu\nu}_{(j)}, \label{4.2}
\end{align}
where, due to the length of their expressions, the respective contributions are explicitly given by Eqs.~\eqref{A.1}.
It should be noticed that the momentum integrals are defined using dimensional regularization, allowing us to define and manipulating them properly.
In particular, we can separate the planar and nonplanar contributions by using the trigonometric relation
$2\sin^{2}\left( \frac{p\times k}{2} \right) = 1 - \cos \left(p\times k\right)$.

As discussed before, we shall consider the low-energy limit of the above self-energy expression \eqref{4.1}. Hence, evaluating the momentum integral with help of Feynman parametrization we find the result for the planar part
 \begin{equation}\label{p-1}
\Pi^{\mu\nu}(p) \Big\vert_{\text{p}}
= \frac{11ig^2}{6\pi}m \Big(\eta^{\mu\nu} - \frac{p^\mu p^\nu}{p^2}\Big),
\end{equation}
in turn, the highly noncommutative limit of the non-planar part, remember that $\tilde{p}$ is kept finite, yields
 \begin{equation}\label{p-2}
\Pi^{\mu\nu}(p) \Big\vert_{\text{n.p}}
= -\frac{i g^2}{32\pi}m\Big(\eta^{\mu\nu}- \frac{p^\mu p^\nu}{p^2}\Big)\Big[42+ \frac{20}{m^2 \tilde{p}^2}  - \frac{11}{m|\tilde{p}|} - \frac{416 m|\tilde{p}|}{9}+\cdots\Big]+\mathcal{O}\Big(\frac{m^4 |\tilde{p}|}{p^2}\Big).
 \end{equation}
As expected the polarization tensor has a parity even structure.
Hence, writing by convenience $\Pi^{\mu\nu}(p)=\Big(\eta^{\mu\nu}- \frac{p^\mu p^\nu}{p^2}\Big)\Pi(p)$, the complete contribution for the one-loop scalar polarization reads
 \begin{equation}\label{p-3}
\Pi (p)=\frac{i g^2 m}{96\pi}\Big[ 50- \frac{60}{m^2 \tilde{p}^2} + \frac{33}{m|\tilde{p}|} + \frac{416 m|\tilde{p}|}{3} +\cdots\Big]+ \mathcal{O}\Big(\frac{m^4 |\tilde{p}|}{p^2}\Big).
 \end{equation}

It is worth noticing that taking the limit $m \rightarrow 0$ does not render a vanishing expression for the scalar polarization, which shows the presence of a UV/IR mixing effect, in contrast with the Maxwell action, which is a free theory.
On the other hand, for the general case where $m \neq 0$, we see that the same terms in Eq.~\eqref{p-3} exhibit the UV/IR mixing effect that does not correspond to any counterpart term in commutative Maxwell theory, and shows that the theory is not infrared finite.

The one-loop polarization tensor \eqref{p-3} allows us to determine the constant $Z_3$ in order to establish the renormalized mass and coupling.
With this result we clearly see that the highly noncommutative limit is interesting because it gives a simpler form for the polarization tensor, allowing thus to easily establish a connection with interesting physical discussion as we will see below.

%%%%%%%%%%%%%%%%%%%%%%%%% %%%%%%%%%%%%%%%%%%%%%%%%%
  %%%%%%%%%%%%%%%%%%%%%%%%% %%%%%%%%%%%%%%%%%%%%%%%%%
 \subsection{One-loop $\left\langle \phi \phi\right\rangle$ self-energy}

 \begin{figure}[t]
\vspace{-1.5cm}
\includegraphics{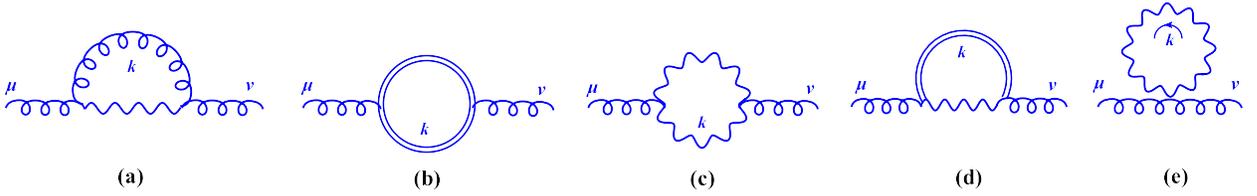}
 \centering
 \caption{Graphs (a) to (e) represent the one-loop contributions to the self-energy of the charged vector field. }
\label{fig2}
\end{figure}

 In the case of the charged vector field self-energy we have five graphs contributing at one-loop order; these are depicted in Fig.~\ref{fig2}.
Once again, all the contributions can easily be constructed with help of the Feynman rules. Since they all have a similar structure, one can write the complete one-loop self energy into a simple form as
\begin{align}
\Lambda^{\mu\nu}(p)=2g^{2} \int \frac{d^{d}k}{(2\pi)^{d}} \frac{M^{\mu\nu}}{[(p+k)^{2}-m^{2}][k^{2}-m^{2}](p+k)^{2}k^{2}} \sin^{2}\big( \frac{p \times k}{2}\big), \label{eq4.4}
\end{align}
where all the contributions were summed into a tensor structure in the numerator
\begin{align}
M^{\mu\nu}=M^{\mu\nu}_{(a)}-2M^{\mu\nu}_{(e)} -m^{2}\big(M^{\mu\nu}_{(b)}+M^{\mu\nu}_{(c)}+M^{\mu\nu}_{(d)}\big),\label{eq4.5}
\end{align}
where the expression for each one of the contributions is explicitly written in Eqs.~\eqref{A.2} of the Appendix \ref{secA}.

The momentum integration can be computed straightforwardly with using the Feynman parametrization. But in order to discuss interesting physical situations we reserve ourselves to the low-energy limit.
In this case, the planar part of the self-energy reads
\begin{align}
\Lambda^{\mu\nu}(p )\Big\vert_{\rm p}
=\frac{17i}{12}mg^{2}\Big(\eta^{\mu\nu}-\frac{p^\mu p^\nu}{p^2}\Big),
\end{align}
while the non-planar part is simply given by
\begin{align}
\Lambda^{\mu\nu}(p)\Big\vert_{\rm n.p}=- \frac{17i}{12\pi}mg^{2}\Big(\eta^{\mu\nu}- \frac{p^\mu p^\nu}{p^2}\Big) e^{-m |\tilde{p}|}.
\end{align}

Hence, writing by convenience $\Lambda^{\mu\nu}(p)=\big(\eta^{\mu\nu}-\frac{p^\mu p^\nu}{p^2}\big)\Lambda(p)$, the complete one-loop contribution to the scalar self-energy function $\Lambda(p)$ at the low-energy limit is \begin{align}
\Lambda(p)=\frac{17i}{12\pi}mg^{2}\left(1-e^{-m|\tilde{p}|}\right)\approx \frac{17i}{12\pi}m^2 g^{2} |\tilde{p}| . \label{eq4.7}
\end{align}
In contrast with Eq.~\eqref{p-3}, we see that Eq.~\eqref{eq4.7}
at the case where $m \neq 0$ and $\theta \neq 0$, presents no UV/IR mixing effect, showing that this sector is infrared finite.
As expected this self-energy function vanishes when $m \rightarrow 0$.
Moreover, we shall use Eq.~\eqref{eq4.7} to compute $Z_2$ and then determine the renormalized mass and coupling.

%

 %%%%%%%%%%%%%%%%%%%%%%%%%%%%%%%%%%%%%%%%%%%%%%%%%
%%%%%%%%%%%%%%%%%%%%%%%%%%%%%%%%%%%%%%%%%%%%%%%%%%
  \subsection{One-loop $\left\langle A \phi\right\rangle$ self-energy}

 \begin{figure}[t]
\vspace{-3cm}
\includegraphics[width=16cm,height=6.5cm]{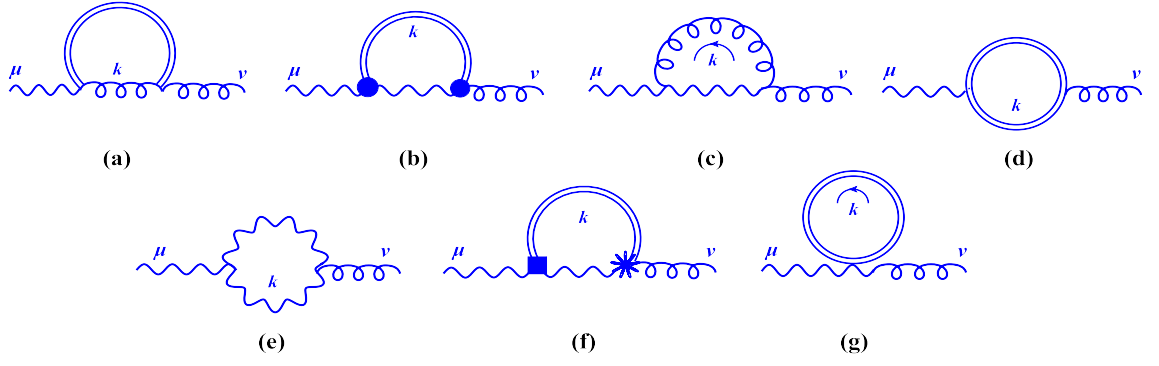}
\centering
\caption{Graphs (a) to (g) represent the one-loop contributions to the self-energy to the mixed propagator $\left\langle A \phi\right\rangle$. }
\label{fig3}
\end{figure}

In the case of the one-loop correction to the mixed propagator $\left\langle A \phi\right\rangle$ we have eight graphs contributing; these are shown in Fig.~\ref{fig3}.
By making use of the Feynman rules these contributions can be cast into a single suitable form. Once again, all these contributions have a similar structure, allowing one to write the complete one-loop self energy as
\begin{align}
\Xi^{\mu\nu}(p) = -2img^{2} \int \frac{d^{d}k}{(2\pi)^{d}} \frac{R^{\mu\nu}}{[(p+k)^{2}-m^{2}][k^{2}-m^{2}](p+k)^{2}k^{2}} \sin^{2}\big(\frac{p \times k}{2}\big),\label{eq4.8}
\end{align}
with the tensor structure conveniently summarized as
\begin{align}
R^{\mu\nu}=R^{\mu\nu}_{(a)} - m^{2}\big(R^{\mu\nu}_{(b)}+R^{\mu\nu}_{(d)}\big) +R^{\mu\nu}_{(c)}+R^{\mu\nu}_{(e,1)}+R^{\mu\nu}_{(e,2)}+R^{\mu\nu}_{(f)}+2R^{\mu\nu}_{(g)},  \label{eq4.9}
\end{align}
where the explicit lengthy expressions $R^{\mu\nu}_{(.)}$ are presented in Eqs.~\eqref{A.3} of the Appendix \ref{secA}.
After computing separately the planar and non-planar parts, we find that the resulting expression reads
\begin{equation}
\Xi (p)=-\frac{g^{2}}{32\pi}\Big[\frac{39}{5}+\frac{66}{m^{2}\tilde{p}^{2}}-\frac{77}{6} m\left|\tilde{p}\right|\Big],  \label{eq4.12}
\end{equation}
where we have made use of the relation $\Xi_{\mu\nu}(p)=\epsilon_{\mu\nu\sigma}p^{\sigma}\Xi(p)$ by simplicity.

With \eqref{eq4.12} we can already proceed to compute the renormalization constant $Z_m$.
It is not difficult to see that for the general case where $m \neq 0$ and $\theta \neq 0$, we see that Eq.~\eqref{eq4.12} displays the UV/IR mixing effect that does not correspond to any counterpart
term in commutative theory, showing thus that the theory is not infrared finite.

Although we can already determine the renormalized mass after determining the constants $Z_3$, $Z_2$ and $Z_m$, we are left to compute the renormalization constants associated with the ghost fields $\left\langle \overline{c}c\right\rangle$ and gauge--ghost vertex $\left\langle A\overline{c}c\right\rangle$ so that the renormalization of the gauge coupling is also established.

   %%%%%%%%%%%%%%%%%%%%%%%%% %%%%%%%%%%%%%%%%%%%%%%%%%
  %%%%%%%%%%%%%%%%%%%%%%%%% %%%%%%%%%%%%%%%%%%%%%%%%%
   \subsection{One-loop $\left\langle \overline{c}c\right\rangle$ self-energy}

\begin{figure}[t]
\vspace{-1cm}
\includegraphics[width=7cm,height=2.3cm]{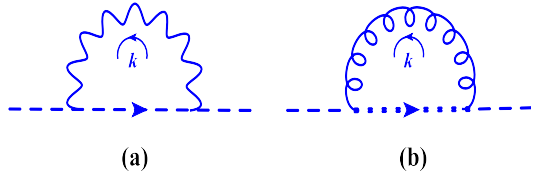}
 \centering
 \caption{Graphs (a) and (b) represent the one-loop contribution to the self-energy to the ghost propagator $\left\langle \bar{c} c\right\rangle$. }
\label{fig4}
\end{figure}

In order to determine the constant $\tilde{Z}_3$ we shall now proceed to the computation of the one-loop self-energy related with the ghost fields. The relevant graphs are shown in Fig.~\ref{fig4}.
These contributions can be written down with Feynman rules as the following
\begin{align}\label{eq4.15}
\mathcal{G}(p) =-8 g^{2}\int\frac{d^{d}k}{(2\pi)^{d}}\frac{\left(p+k\right)^{\mu}p^{\nu}\left(k^{2}\eta_{\nu\mu}-k_{\nu}k_{\mu}\right)}
{k^{2}(k^{2}-m^{2})(p+k)^{2}}\sin^{2}\big(\frac{p\times k}{2}\big).
\end{align}
We can readily compute the planar contribution as
\begin{align}
\mathcal{G}(p)\Bigr|_{\textrm{p}}=-\frac{ig^{2}}{2\pi}p^{2}\int_{0}^{1}dz\int_{0}^{1-z}dy\frac{1}{\sqrt{\Delta}},
\end{align}
where $\Delta=ym^{2}-z\left(1-z\right)p^{2}$, while the computation of the non-planar part is rather complicated and reads
\begin{align}
\mathcal{G}(p)\Bigr|_{\textrm{n.p}}  =\frac{ig^{2}}{2\pi}p^{2}\int_{0}^{1}dz\int_{0}^{1-z}dy\frac{1}{\sqrt{\Delta}}
\Big(1+\frac{\sqrt{\Delta}\left|\tilde{p}\right|}{2}\Big)e^{-\sqrt{\Delta}\left|\tilde{p}\right|}.
\end{align}
In accordance with the previous analysis we consider the full contribution in the low-energy limit, i.e. $p^{2}/m^{2}\rightarrow0$,
so that we obtain a simple result
\begin{align} \label{eq4.16}
\mathcal{G}(p) \approx-\frac{ig^{2}}{4\pi}p^{2}\left|\tilde{p}\right|\int_{0}^{1}dz\int_{0}^{1-z}dy\left[1+\sqrt{\Delta}
\left|\tilde{p}\right|+\cdots\right]\approx-\frac{ig^{2}}{8\pi}p^{2}\left|\tilde{p}\right|.
\end{align}
Hence the leading contribution for the ghost self-energy given by \eqref{eq4.16} is finite and does not presents the UV/IR mixing effect. This simple expression implies into a straightforward computation of the respective renormalization constant.

     %%%%%%%%%%%%%%%%%%%%%%%%% %%%%%%%%%%%%%%%%%%%%%%%%%
  %%%%%%%%%%%%%%%%%%%%%%%%% %%%%%%%%%%%%%%%%%%%%%%%%%
  \subsection{One-loop $\left\langle A\overline{c}c\right\rangle$ vertex}

\begin{figure}[t]
\vspace{-1cm}
\includegraphics[width=9cm,height=4cm]{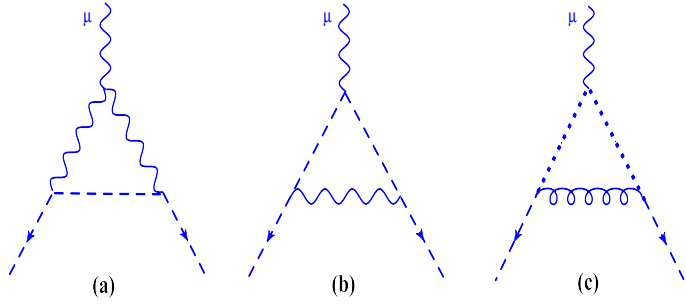}
 \centering
 \caption{Graphs (a) to (c) represent the one-loop contribution to the vertex $\left\langle \bar{c} c A\right\rangle$. }
\label{fig5}
\end{figure}

 At last, we proceed to compute the one-loop correction to the $\left\langle A\overline{c}c\right\rangle$ vertex function. There are three diagrams contributing at this order, these are shown at Fig.~\ref{fig5}.
The first contribution can be expressed as
\begin{align}\label{eq4.20}
\Theta_{(a)}^{\mu}(p,q,q-p) =-8ig^{3}\int\frac{d^{d}k}{(2\pi)^{d}}~\frac{\sin\big(\frac{p\times k}{2}\big)\sin\big(\frac{\left(q-p\right)\times k}{2}\big)\sin\big(\frac{q\times(p+k)}{2}\big)}{k^{2}\left(k^{2}-m^{2}\right)(q-p-k)^{2} \left(k+p\right)^{2}\left(\left(k+p\right)^{2}-m^{2}\right)}~{\cal{N}}_{(a)}^{\mu},
\end{align}
with the numerator written as
\begin{align}
{\cal{N}}_{(a)}^{\mu} &=\left(q-p-k\right)^{\alpha}q^{\beta} \Big[(2p+k)^{\nu}\eta^{\mu\rho}+\left(2k+p\right)^{\mu}\eta^{\nu\rho}+\left(k-p\right)^{\rho}\eta^{\mu\nu}\Big] \nonumber\\
&\times  \left(\eta_{\nu\alpha}k^{2}-k_{\nu}k_{\alpha}\right)\left(\eta_{\beta\rho}
\left(k+p\right)^{2}-\left(k+p\right)_{\beta}\left(k+p\right)_{\rho}\right),
\end{align}
while one can easily show that the diagrams (b) and (c) give the same contribution,
\begin{align} \label{eq4.21}
\Theta_{(b)}^{\mu}(p,q,q-p)=-8ig^{3}\int\frac{d^{d}k}{(2\pi)^{d}}~ \frac{\sin\big(\frac{p\times\left(k-q\right)}{2}\big)\sin\big(\frac{\left(q-p\right)\times k}{2}\big)\sin\big(\frac{q\times k}{2}\big)}{k^{2}\left(k^{2}-m^{2}\right)(q-p-k)^{2} \left(q-k\right)^{2}}~{\cal{N}}_{(b)}^{\mu},
\end{align}
where
\begin{align}
{\cal{N}}_{(b)}^{\mu}=\left(q-k\right)^{\mu}\left(q-p-k\right)^{\nu}\left(\eta_{\nu\rho}k^{2}-k_{\nu}k_{\rho}\right)q^{\rho}
\end{align}
In this sense, we have that the full contribution is given by
\[
\Theta^{\mu}(p,q,q-p)=\Theta_{(a)}^{\mu}(p,q,q-p)+2\Theta_{(b)}^{\mu}(p,q,q-p)
\].
However, the computation of the expressions \eqref{eq4.20} and \eqref{eq4.21} involve complicated manipulations to separate the planar and non-planar parts; moreover, such expressions vanish in the highly noncommutative limit. The full expression for the diagram (a) can then be written as
\begin{align} \label{eq4.22}
\Theta_{(a)}^{\mu}(p,q,q-p)\approx-\frac{g^{3}}{16\pi}q^{\mu}q^{2}\sin\big(\frac{p\times q}{2}\big)F\left(q,p,m\right),
\end{align}
whereas the diagram (b) reads
\begin{align}
\Theta_{(b)}^{\mu}(p,q,q-p) \label{eq4.23}
 \approx-\frac{g^{3}}{16\pi}q^{\mu}q^{2}\sin\big(\frac{p\times q}{2}\big)G\left(q,p,m\right),
\end{align}
where $F\left(q,p,m\right)$ and $G\left(q,p,m\right)$ are a finite, but complicated, function of the external momenta and Feynman parameters.

In this sense, the one-loop contribution is expressed as
\begin{equation}\label{eq4.24}
\Theta^{\mu}(p,q,q-p)\approx-\frac{g^{3}}{16\pi}q^{\mu}q^{2}\sin\big(\frac{p\times q}{2}\big)H\left(q,p,m\right),
\end{equation}
where we have introduced by simplicity $H\left(q,p,m\right)=F\left(q,p,m\right)+2G\left(q,p,m\right)$.
Notice, however, since the contribution \eqref{eq4.24} is proportional to the external momentum, we see that it vanishes rapidly when $p^{2}/m^{2}\rightarrow0$. In this sense, although finite, the one-loop contribution $\left\langle A\overline{c}c\right\rangle$ vertex function is zero in the highly noncommutative limit, which implies that the respective renormalization constant is equal to unit, i.e. $\tilde{Z}_{3}^{\rm gh}=1$.

  %%%%%%%%%%%%%%%%%%%%%%%%% %%%%%%%%%%%%%%%%%%%%%%%%%
  %%%%%%%%%%%%%%%%%%%%%%%%% %%%%%%%%%%%%%%%%%%%%%%%%%
 \section{Renormalized mass and charge}
\label{sec5}

After computing explicitly the one-loop corrections to the 1PI functions of interest, we can now proceed in determining the effect of such corrections on the behavior of the physical mass and charge.
By writing the renormalization constant in terms of a counterterm, $Z_i =1 + \delta_{Z_i}$, the renormalized mass $m=\frac{\sqrt{Z_{2} Z_{3}} }{Z_m} m_0$ reads
\begin{equation}
m^{2}_{\rm ren}=m^2 +m ^2\big( \delta_{Z_2}+ \delta_{Z_3} -2\delta_{Z_m}\big), \label{eq5.1}
\end{equation}
where each counterterm is evaluated under the conditions
\begin{equation}
i\delta_{Z_3} =  \frac{1}{p^2}\Pi(p)\Big\vert_{ p^2 = m^2}, \quad
i \delta_{Z_2} = \frac{1}{p^2}\Lambda(p)\Big\vert_{ p^2 = m^2}, \quad
\delta_{\tilde{Z}_m} = - \frac{1}{m} \Xi(p)\Big\vert_{ p^2 = m^2}. \label{eq5.3}
\end{equation}
Due to our interest, we shall determine the above counterterms in the low-energy limit, so that we can make use of the previous results of the radiative corrections.
%To compute the $\left\langle AA\right\rangle$ counter-term we resort to the result \eqref{p-3}, in this case we find
%\begin{equation}
%\delta_{Z_3} = \frac{ g^2}{96\pi m}   \Big[ 50- \frac{60}{m^2 \tilde{p}^2}  + \frac{33}{m|\tilde{p}|} + \frac{416 m|\tilde{p}|}{3} +\cdots\Big]+ \mathcal{O}\Big(\frac{m^3 |\tilde{p}|}{p^2}\Big),
%\end{equation}
%while for the $\left\langle \phi \phi \right\rangle$ counter-term we make use of \eqref{eq4.7}, so that it yields
%\begin{equation}
%\delta_{Z_2} = \frac{17g^{2}}{12\pi} |\tilde{p}|,
%\end{equation}
%in turn, the mass counter-term, related with the mixed function $\left\langle A \phi \right\rangle$, is determined from Eq.~\eqref{eq4.12} as
%\begin{equation}
%\delta_{Z_m} = \frac{g^{2}}{32\pi m}\Big[\frac{39}{5}+\frac{66}{m^{2}\tilde{p}^{2}}-\frac{77}{6} m\left|\tilde{p}\right| \Big] .
%\end{equation}
Notice that with the renormalized mass \eqref{eq5.1} we can determine the dispersion relation for the fields $A_\mu$ and $\phi_\mu$, $p^2 = m^2_{\rm ren}$, which is corrected by
\begin{equation}
\omega^2_p =\vec{p}^2 + m^2 +m^2 \big( \delta_{Z_2}+ \delta_{Z_3} -2\delta_{Z_m}\big) .
\end{equation}
Hence, making use of the above expressions for the counterterms, the one-loop physical dispersion relation is explicitly written as
\begin{equation}
\omega^2_p =\vec{p}^2 + m^2 +\frac{m g^{2}}{96\pi } \Big(\frac{16}{5}- \frac{456}{m^2 \tilde{p}^2} + \frac{33}{m|\tilde{p}|}   \Big) +\mathcal{O}\Big( \frac{g^2 m^2 |\tilde{p}|}{p^2}\Big). \label{eq5.2}
\end{equation}
From this expression one can define a physical mass $m_{\rm phys}^2 = m^2 +\frac{m g^{2}}{30\pi }$ due to the one-loop effects of the radiative corrections. Notice that this mass shift arises from the planar NC effect. Moreover, from Eq.~\eqref{eq5.2} we can see the presence of a severe UV/IR instability in the NC momentum, owing to the nonplanar NC effects, affecting the propagation of the gauge and vector fields, this instability originates from Eqs.~\eqref{p-3} and \eqref{eq4.12}, i.e., those corrections involving the gauge field.

It is notable that the form of the terms representing UV/IR instability in \eqref{eq5.2} is quite different from that of NC Maxwell-Chern-Simons (M-CS) theory \cite{Ghasemkhani:2015tqu} and also of NC QED$_{4}$ \cite{Matusis:2000jf}. Here in the NC Jackiw-Pi model, there are strong instabilities due to $\frac{1}{\tilde{p}^{2}}$ and a soft one by $\frac{1}{|\tilde{p}|}$, while in the NC QED$_{4}$ we have again a $\frac{1}{\tilde{p}^{2}}$ instability, and in the NC M-CS theory a soft $\frac{1}{|\tilde{p}|}$ instability. We note that the NC Jackiw-Pi model has a severe instability when compared to the NC M-CS theory. Furthermore, in contrast to \eqref{eq5.2}, where we have a shift on the mass coming from the planar NC effect, we do not have any correction to the mass in the case of both NC M-CS and NC QED$_{4}$.

In regard to the renormalization of the gauge coupling, we should first recall that due to the vanishing behavior of \eqref{eq4.24} in $p^{2}/m^{2}\rightarrow0$, it is easy to conclude that $\tilde{Z}_{3}^{\rm gh}=1$.
In this case, the renormalization constant related to the gauge coupling is simplified to $Z_{g}=Z_{3}^{1/2}\tilde{Z}_{3}$. In terms of its counterterms we rewrite the constant as
\begin{equation}
Z_{g}= 1 + \frac{1}{2}\delta_{Z_3} + \delta_{\tilde{Z}_3},
\end{equation}
where the gauge-field related counterterm is given by \eqref{eq5.3}, while the ghost counterterm can be determined by the condition
\begin{equation}
\delta_{\tilde{Z}_3} = - \frac{1}{p^2}\mathcal{G}(p)\Big\vert_{ p^2 = 0}.
\end{equation}
%Hence, from \eqref{eq4.16} we find that
%\begin{equation}
%\delta_{\tilde{Z}_3} =  \frac{g^{2}}{8\pi} \left|\tilde{p}\right|.
%\end{equation}
Finally, we realize that the physical behavior of the coupling constant is
\begin{equation}
g_{\rm ren}= g+\frac{25 g^3}{48\pi m}.
\end{equation}
We observe that the coupling constant gets renormalized by planar NC effects, which are absent in the Abelian Jackiw-Pi model, being a free theory.
It is worth mentioning that the coupling constant $g$ in a $(2+1)$ spacetime is dimensionful and hence in order to study the physical behavior of the theory under a change of the energy scale, it is useful to introduce a dimensionless coupling constant to get a more physical information. To this end, it is convenient to rewrite the renormalized mass and charge as below
\begin{equation}
 g_{\rm ren}=g\big(1+a\lambda\big),\quad m^2_{\rm ren}=m^2\big(1+b\lambda\big),
 \label{eq5.8}
\end{equation}
where $a=\frac{25}{48\pi}$, $b=\frac{1}{30\pi}$ and $\lambda=\frac{g^2}{m}$. From \eqref{eq5.8}, we realize that the ratio $\lambda=\frac{g^2}{m}$ is in fact the dimensionless expansion parameter in such dimensionality. Therefore, the physical behavior of this dimensionless coupling constant can be expressed by
\begin{equation}
\lambda_{\rm ren}=\lambda\big(1+c\lambda\big),
\end{equation}
where $c=2a -\frac{b}{2}$ and we have also neglected higher order terms in $\lambda$ inside the parentheses.
Indeed, it is expected to find such a relation for this model in three dimensions, since the mass dimension of $g$ is $\frac{1}{2}$ and hence the only dimensionless coupling constant is described as $\lambda=\frac{g^2}{m}$.
We can make this discussion clearer by making the scale changes $g A_\mu \rightarrow \tilde{A}_\mu$ and $g \phi_\mu \rightarrow \tilde{\phi}_\mu$ in the Lagrangian \eqref{eq:1.0}, so that it reads
\begin{equation}
\mathcal{L}=-\frac{1}{4g^2}\tilde{F}_{\mu\nu}\star \tilde{F}^{\mu\nu}-\frac{1}{4g^2}\tilde{G}_{\mu\nu} \star \tilde{G}^{\mu\nu}+\frac{1}{2\lambda}\epsilon^{\mu\nu\alpha}\tilde{F}_{\mu\nu}\star\tilde{\phi}_{\alpha}.
\end{equation}
We notice two interesting limiting cases:
\begin{itemize}
\item In the limit $g^2 \rightarrow \infty$ and $m \rightarrow  \infty$, we keep the ratio $\lambda$ finite, in this case only the mixing term survives.

\item On the other hand, when $\lambda \rightarrow \infty$, i.e., $m \rightarrow 0$ keeping $g^2$ finite, the mixing term vanishes and we have massless fields, where they couple solely through the minimal coupling in $\tilde{G}_{\mu\nu}$.
\end{itemize}
Since the theory is UV finite, further information would only be achievable through a nonperturbative approach for the beta function.

%%%%%%%%%%%%%%%%%%%%%%%%%%%%%%%%%%%%%%%%%%%%%%%%%%%%%%%%%%%%%%%%%%%%%%%
%%%%%%%%%%%%%%%%%%%%%%%%%%%%%%%%%%%%%%%%%%%%%%%%%%%%%%%%%%%%%%%%%%%%%%%
\section{Final remarks}
\label{conc}

In this paper, we have studied the physical aspects of the noncommutative Jackiw-Pi model. In order to establish the behavior of the renormalized parameters within the model, we have proceeded with the computation of the necessary one-loop corrections, although all of them are UV finite, some of these functions present UV/IR instabilities in their expressions, which in turn imply instabilities in the propagation of the gauge field.

We started by reviewing the main aspects concerning the gauge structure of the Jackiw-Pi model, with particular interest in the BRST transformation, allowing thus a consistent construction of a BRST invariant noncommutative Jackiw-Pi model. Establishing the BRST structure of the NC Jackiw-Pi model, we then proceeded to study the one-loop renormalization of this model, writing the renormalized mass and gauge coupling. In order to compute these renormalized quantities, we compute the necessary one-loop 1PI functions, although all corrections were UV finite, the $\langle AA\rangle$ and $\langle A\phi\rangle$ self-energy expressions displayed UV/IR instabilities, which are then reflected in the physical behavior of the physical renormalized quantities.
It is worth noticing that though the presence of such instabilities, the tree level parity and gauge invariance of the NC Jackiw-Pi model were preserved at the one-loop quantum level.

Furthermore, it is worth to call attention to the fact that the one-loop analysis at the low-energy limit of the NC Jackiw-Pi model exhibits a finite shift for the gauge field mass as $m^{2}_{\rm phys}=m^{2}+\frac{mg^{2}}{30\pi}$,  while there is no any one-loop correction to the mass at the low-energy limit in the NC Maxwell-Chern-Simons model \cite{Ghasemkhani:2015tqu}.
The UV/IR instabilities discussed here when the limit $|\theta_{\mu\nu}| \rightarrow 0$ is taken on the one-loop self-energy functions are clearly engendered by quantum effects, since this limit and integration sign do not commute. Moreover, these instabilities are a shared feature of both parity even Jackiw-Pi model and Maxwell-Chern-Simons model \cite{Ghasemkhani:2015tqu} when described in the noncommutative framework at a quantum level, while it seems that these instabilities are absent when supersymmetry is added, as it is the case of the NC ABJM model \cite{Martin:2017nhg}.
Due to these results, one can naively think that the addition of invariance under discrete symmetries is not sufficient to remove those undesired instabilities, but rather an enlarged continuous symmetry invariance such as supersymmetry can achieve a consistent and true finite result for NC field theories.

%%%%%%%%%%%%%%%%%%%%%%%%%%%%%%%%%%%%%%%%%%%%%%%%%%%%%%%%%%%%%%%%%%%%%%%
%%%%%%%%%%%%%%%%%%%%%%%%%%%%%%%%%%%%%%%%%%%%%%%%%%%%%%%%%%%%%%%%%%%%%%%
 \subsection*{Acknowledgements}
The authors are grateful to M. M. Sheikh-Jabbari for his valuable
remarks and fruitful discussions. Also, we would like to thank C. P.
Martin for his comments on the manuscript. R.B. acknowledges partial
support from Conselho Nacional de Desenvolvimento Cient\'{i}fico e
Tecnol\'{o}gico (CNPq Project No. 304241/2016-4) and Funda\c{c}\~{a}o de Amparo \`a
Pesquisa do Estado de Minas Gerais (FAPEMIG Project No. APQ-01142-17).

\appendix
%%%%%%%%%%%%%%%%%%%%%%%%
%%%%%%%%%%%%%%%%%%%%%%%%%%%%
\section{Tensor structures}
\label{secA}

In order to avoid lengthy expressions along the main text, we present for completeness some important tensor structures from the self-energy functions in the Sec.~\ref{sec4}.
First, from the one-loop correction for the gauge field $\left\langle AA\right\rangle$ \eqref{4.1} we have
\begin{align*}
N^{\mu\nu}_{(a)} &= \left[(p-k)^{\alpha}\eta^{\mu\rho} + (p+2k)^{\mu}\eta^{\rho\alpha}-(2p+k)^{\rho}\eta^{\alpha\mu}\right]\left[ (p+k)^{2}\eta_{\alpha\beta} - (p+k)_{\alpha}(p+k)_{\beta}\right]\nonumber\\
&\times \left[ -(2p+k)^{\sigma}\eta^{\nu\beta} + (p+2k)^{\nu}\eta^{\beta\sigma} + (p-k)^{\beta}\eta^{\sigma\nu}\right]\left[ k^{2}\eta_{\sigma\rho} - k_{\sigma}k_{\rho}\right],
\nonumber\\
\vspace{7mm}
N^{\mu\nu}_{(b)} &=[(p+k)^{2}-m^{2}](p+k)^{2}[2k^{2}\eta^{\mu\nu} + 2k^{\mu}k^{\nu}],
\nonumber\\
N^{\mu\nu}_{(c)} &= \left[k^{\alpha}\eta^{\mu\rho} - (p+2k)^{\mu}\eta^{\rho\alpha}+(p+k)^{\rho}\eta^{\alpha\mu}\right]\left[ (p+k)^{2}\eta_{\alpha\beta} - (p+k)_{\alpha}(p+k)_{\beta}\right]\nonumber\\
&\times \left[ (p+k)^{\sigma}\eta^{\nu\beta} - (p+2k)^{\nu}\eta^{\beta\sigma} +k^{\beta}\eta^{\sigma\nu}\right]\left[ k^{2}\eta_{\sigma\rho} - k_{\sigma}k_{\rho}\right],
\nonumber\\
N^{\mu\nu}_{(d)} &=[(p+k)^{2}-m^{2}](p+k)^{2}[-k^{2}\eta^{\mu\nu} - k^{\mu}k^{\nu}],
\nonumber\\
N^{\mu\nu}_{(e)} &= \epsilon^{\mu\rho\alpha}\epsilon_{\alpha\beta\lambda}\epsilon^{\nu\beta\sigma}\epsilon_{\sigma\rho\chi}(p+k)^{\lambda}k^{\chi} ,  \nonumber\\
N^{\mu\nu}_{(f)} &= \epsilon^{\mu\rho\alpha}\epsilon^{\nu\beta\sigma} \left[ (p+k)^{2} \eta_{\alpha\beta} - (p+k)_{\alpha}(p+k)_{\beta}\right]\left[k^{2}\eta_{\sigma\rho} - k_{\sigma}k_{\rho} \right],
\nonumber\\
N^{\mu\nu}_{(g)} &= \epsilon_{\alpha\beta\lambda}\epsilon^{\nu\beta\sigma}(p+k)^{\lambda}[k^{2}\eta_{\sigma\rho} - k_{\sigma}k_{\rho}]\big[ (p-k)^{\alpha}\eta^{\mu\rho}+ (p+2k)^{\mu}\eta^{\rho\alpha} - (2p+k)^{\rho}\eta^{\alpha\mu}\big], \nonumber\\
N^{\mu\nu}_{(h.1)} &=\epsilon_{\alpha\beta\lambda}\epsilon_{\sigma\rho\chi}(p+k)^{\lambda}k^{\chi} \big[ (p-k)^{\alpha}\eta^{\mu\rho} + (p+2k)^{\mu}\eta^{\rho\alpha} - (2p+k)^{\rho}\eta^{\alpha\mu} \big]\nonumber\\
&\times [(p+k)^{\sigma}\eta^{\nu\beta}- (p+2k)^{\nu}\eta^{\beta\sigma} + k^{\beta}\eta^{\sigma\nu}\big],
\end{align*}
%%%%
\begin{align}
N^{\mu\nu}_{(h.2)} &= \epsilon_{\alpha\beta\lambda}\epsilon_{\sigma\rho\chi}(p+k)^{\lambda}k^{\chi}\big[ k^{\alpha}\eta^{\mu\rho} - (p+2k)^{\mu}\eta^{\rho\alpha} + (p+k)^{\rho}\eta^{\alpha\mu}\big]\nonumber\\
&\times \big[-(2p+k)^{\sigma}\eta^{\nu\beta}+ (p+2k)^{\nu}\eta^{\beta\sigma} + (p-k)^{\beta}\eta^{\sigma\nu}\big],
\nonumber\\
N^{\mu\nu}_{(i)} &= \epsilon_{\alpha\beta\lambda}\epsilon^{\nu\beta\sigma}(p+k)^{\lambda}\big[ k^{\alpha}\eta^{\mu\rho} - (p+2k)^{\mu}\eta^{\rho\alpha} + (p+k)^{\rho}\eta^{\alpha\mu}\big]\big[k^{2}\eta_{\sigma\rho} - k_{\sigma}k_{\rho}\big], \nonumber\\
N^{\mu\nu}_{(j)} &= [(p+k)^{2} -m^{2}][k^{2}-m^{2}] k^{\mu}(p+k)^{\nu}.  \label{A.1}
\end{align}
 Next, for the vector field $\left\langle \phi \phi \right\rangle$ contributions, the relevant tensor part from Eq.~\eqref{eq4.4} reads
 expression
\begin{align}
M^{\mu\nu}_{(a)} &= \left[k^{\alpha}\eta^{\mu\rho} - (p+2k)^{\mu}\eta^{\rho\alpha}+(p+k)^{\rho}\eta^{\alpha\mu}\right]\left[ (p+k)^{2}\eta_{\alpha\beta} - (p+k)_{\alpha}(p+k)_{\beta}\right]\nonumber\\
&\times  \left[ (p+k)^{\sigma}\eta^{\nu\beta} - (p+2k)^{\nu}\eta^{\beta\sigma} +k^{\beta}\eta^{\sigma\nu}\right]\left[ k^{2}\eta_{\sigma\rho} - k_{\sigma}k_{\rho}\right],\nonumber\\
M^{\mu\nu}_{(b)} &= \epsilon_{\alpha\beta\lambda}\epsilon_{\sigma\rho\chi}(p+k)^{\lambda}k^{\chi}\big[k^{\alpha}\eta^{\mu\rho} - (p+2k)^{\mu}\eta^{\rho\alpha} + (p+k)^{\rho}\eta^{\alpha\mu}\big]\nonumber\\
&\times [(p+k)^{\sigma}\eta^{\nu\beta} - (p+2k)^{\nu}\eta^{\beta\sigma} + k^{\beta}\eta^{\sigma\nu}\big] ,\nonumber\\
M^{\mu\nu}_{(c)} &= \epsilon^{\mu\rho\alpha}\epsilon^{\nu\beta\sigma} \left[ (p+k)^{2} \eta_{\alpha\beta} - (p+k)_{\alpha}(p+k)_{\beta}\right]\left[k^{2}\eta_{\sigma\rho} - k_{\sigma}k_{\rho} \right], \nonumber\\
M^{\mu\nu}_{(d)} &= \epsilon^{\mu\rho\alpha}\epsilon_{\sigma\rho \chi}k^{\chi} \left[ (p+k)^{2} \eta_{\alpha\beta} - (p+k)_{\alpha}(p+k)_{\beta}\right]\left[\left( p+k\right)^{\sigma}\eta^{\nu\beta}-\left( 2k+p\right)^{\nu}\eta^{\beta\sigma}+k^{\beta}\eta^{\sigma\nu}  \right],\nonumber\\
M^{\mu\nu}_{(e)} &= [(p+k)^{2}-m^{2}](p+k)^{2}[-k^{2}\eta^{\mu\nu} - k^{\mu}k^{\nu}].  \label{A.2}
\end{align}
Also, for the one-loop correction of the mixed propagator $\left\langle A \phi\right\rangle$, we have from \eqref{eq4.8} that
 \begin{align}
R^{\mu\nu}_{(a)} &= \epsilon_{\alpha \beta \lambda} (p+k)^{\lambda} \big[k^{\alpha}\eta^{\mu\rho} - (p+2k)^{\mu}\eta^{\rho\alpha} + (p+k)^{\rho} \eta^{\alpha\mu} \big] \big[k^{2}\eta_{\sigma\rho} - k_{\sigma}k_{\rho} \big] \nonumber\\
&\times \big[(p+k)^{\sigma} \eta^{\nu\beta} - (p+2k)^{\nu}\eta^{\beta\sigma} + k^{\beta}\eta^{\sigma\nu} \big],\nonumber\\
R^{\mu\nu}_{(b)} &= \epsilon^{\mu\rho\alpha}\epsilon^{\nu\beta\sigma}\epsilon_{\sigma\rho\chi} k^{\chi} \big[ (p+k)^{2} \eta_{\alpha\beta} - (p+k)_{\alpha} (p+k)_{\beta} \big], \nonumber\\
R^{\mu\nu}_{(c)} &= \epsilon^{\mu\rho\alpha} \big[ (p+k)^{2} \eta_{\alpha\beta} - (p+k)_{\alpha} (p+k)_{\beta} \big]\big[k^{2}\eta_{\sigma\rho} - k_{\sigma}k_{\rho}\big] \nonumber\\
&\times \big[ (p+k)^{\sigma}\eta^{\nu\beta} - (p+2k)^{\nu}\eta^{\beta\sigma} + k^{\beta} \eta^{\sigma\nu}\big],\nonumber\\
R^{\mu\nu}_{(d)} &= \epsilon^{\mu\rho\alpha} \epsilon_{\alpha\beta\lambda}\epsilon_{\sigma\rho\chi} (p+k)^{\lambda}k^{\chi}\big[(p+k)^{\sigma}\eta^{\nu\beta} - (p+2k)^{\nu}\eta^{\beta\sigma} + k^{\beta}\eta^{\sigma\nu}\big] , \nonumber\\
R^{\mu\nu}_{(e.1)} &= \epsilon^{\nu\beta\sigma}\big[(p-k)^{\alpha}\eta^{\mu\rho} + (p+2k)^{\mu}\eta^{\rho\alpha} - (2p+k)^{\rho}\eta^{\alpha\mu} \big]\nonumber\\
&\times \big[(p+k)^{2}\eta_{\alpha\beta} - (p+k)_{\alpha}(p+k)_{\beta}\big]\big[ k^{2} \eta_{\sigma\rho} - k_{\sigma}k_{\rho} \big], \nonumber\\
R^{\mu\nu}_{(e.2)} &= \epsilon^{\mu\rho\alpha}\big[ - (2p+k)^{\sigma} \eta^{\nu\beta} + (p+2k)^{\nu} \eta^{\beta\sigma} + (p-k)^{\beta}\eta^{\sigma\nu}\big]\nonumber\\
&\times \big[k^{2}\eta_{\sigma\rho} - k_{\sigma}k_{\rho}\big]\big[(p+k)^{2}\eta_{\alpha\beta} - (p+k)_{\alpha}(p+k)_{\beta}\big]  , \nonumber\\
R^{\mu\nu}_{(f)} &= \epsilon_{\sigma\rho\chi}k^{\chi}\big[(p-k)^{\alpha}\eta^{\mu\rho} + (p+2k)^{\mu}\eta^{\rho\alpha} - (2p+k)^{\rho}\eta^{\alpha\mu} \big] \nonumber\\
&\times \big[(p+k)^{2}\eta_{\alpha\beta} - (p+k)_{\alpha}(p+k)_{\beta}\big]\big[ (p+k)^{\sigma}\eta^{\nu\beta} - (p+2k)^{\nu}\eta^{\beta\sigma} + k^{\beta}\eta^{\sigma\nu}\big] ,\nonumber\\
R^{\mu\nu}_{(g)}  &= \epsilon^{\mu\nu\lambda}k_{\lambda}[(p+k)^{2}- m^{2} ] (p+k)^{2}. \label{A.3}
\end{align}

%%%%%%%%%%%%%%%%%%%%%%%%%%%%%%%%%%%%%%%%%%%%%%%%%%%%%%%%%%%%%%%%%%%%%%%%%%%%%%%%%%%%%%%%%%%%%%%%%%%%%%%%%%%%%%%%%%%%%%%%%%%%%%%%%%%%%%%%%%%%%%%%%%%%%%%%%%%%%%%%%%%%%%%%%%%%%%%%%%%%%%%%%%%%%%%%%%%%%%%%%%%%%%%%%%%%%%

\global\long\def\link#1#2{\href{http://eudml.org/#1}{#2}}
 \global\long\def\doi#1#2{\href{http://dx.doi.org/#1}{#2}}
 \global\long\def\arXiv#1#2{\href{http://arxiv.org/abs/#1}{arXiv:#1 [#2]}}
 \global\long\def\arXivOld#1{\href{http://arxiv.org/abs/#1}{arXiv:#1}}

%%%%%%%%%%%%%%%%%%%%%%%%%%%%%%%%%%%%%%%%%%%%%%%%%%%%%%%%%%%%%%%%%%%%%%%%%%%%%%%%%%%%%%%%%%%%%%%%%%%%%%%%%%%


\begin{thebibliography}{99}
%%%%%%%%%%%%%%%%%%%%%%%%%%%%%%%%%%%%%%%%%%%%%%%%%%%%%%%%%%%%%%%%%%%%%%%%%%%%%%%%%%%%%%%%%%%%%%%%%%%%%%%%%%%

%%%%%%%%%%%%%%%%%%%%%%%%%%%%%%%%%%%%%%%%%%%%%%%%%%%%%%
  \bibitem{schwinger1}
  J.~S.~Schwinger,
  \textit{``Gauge invariance and mass,}''
    \doi{10.1103/PhysRev.125.397}{Phys.\ Rev.\  {\bf 125}, 397 (1962)}.
  %%%%%%%%%%%%%%%%%%%%%%%%%%%%%%%%%%%%%%%%%%%%%%%%%%%%%%%
    \bibitem{schwinger2}
  J.~S.~Schwinger,
  ``\textit{Gauge invariance and mass \emph{II},}''
  \doi{10.1103/PhysRev.128.2425}{Phys.\ Rev.\ {\bf 128}, 2425 (1962)}.
%%%%%%%%%%%%%%%%%%%%%%%%%%%%%%%%%%%%%%%%%%%
%%ref-ref-24
\bibitem{jackiw}
S.~Deser, R.~Jackiw and S.~Templeton,
  ``\textit{Topologically Massive Gauge Theories},''
   \doi{10.1016/0003-4916(82)90164-6}{Annals Phys.\  {\bf 140}, (1982) 372};
   {\color{blue}{\bf 185} 406 (1988);}
  \doi{10.1006/aphy.2000.6013}{{\bf 281}, 409 (2000)}.
%%%%%%%%%%%%%%%%%%%%%%%%%%%%%%%%%%%%%%%%%
\bibitem{Marino:2017ckg}
  E.~C.~Marino,
  `` \textit{Quantum Field Theory Approach to Condensed Matter Physics,}''
  Cambridge University Press, 2017.

%%%%%%%%%%%%%%%%%%%%%%%%%%%%%%%%%%%%%%%%%%%%%%%%%%%%%%%%%%%%%%%%%%%%%%%%%%%%%%%%%%%%%%%%%%%%%%%%%%%%%%%%%%%
  \bibitem{Jackiw:1997jga}
  R.~Jackiw and S.~Y.~Pi,
  ``Seeking an even parity mass term for 3-D gauge theory,''
  \doi{10.1016/S0370-2693(97)00520-0}{ Phys.\ Lett.\ B {\bf 403}, 297 (1997)},
       \arXivOld{hep-th/9703226}.

       \bibitem{Jackiw:1997ha}
  R.~Jackiw,
  `` \textit{NonYang-Mills gauge theories,}''
  In *Peniscola 1997, Advanced school on non-perturbative quantum field physics* 201-222
\arXivOld{hep-th/9705028}.
 %%%%%%%%%%%%%%%%%%%%%%%%%%%%%%%%%%%%%%%%%%%%%%%%%%%%%%%%%%%%%%%%%%%%%%%%%%%%%%%%%%%%%%%%%%%%%%%%%%%%%%%%%%%
\bibitem{Henneaux:1997mf}
  M.~Henneaux, V.~E.~R.~Lemes, C.~A.~G.~Sasaki, S.~P.~Sorella, O.~S.~Ventura and L.~C.~Q.~Vilar,
  `` \textit{A No go theorem for the nonAbelian topological mass mechanism in four dimensions,}''
 \doi{10.1016/S0370-2693(97)00984-2}{Phys.\ Lett.\ B {\bf 410}, 195 (1997)},
       \arXivOld{hep-th/9707129}.
    %%%%%%%%%%%%%%%%%%%%%%%%%%%%%%%%%%%%%%%%%%%%%%%%%%%%%%%%%%%%%%%%%%%%%%%%%%%%%%%%%%%%%%%%%%%%%%%%%%%%%%%%%%%
\bibitem{Dayi:1997in}
  O.~F.~Dayi,
  `` \textit{Hamiltonian formulation of Jackiw-Pi three-dimensional gauge theories,}''
\doi{10.1142/S0217732398002072}{   Mod.\ Phys.\ Lett.\ A {\bf 13}, 1969 (1998)},
       \arXivOld{hep-th/9711079}.
  %%%%%%%%%%%%%%%%%%%%%%%%%%%%%%%%%%%%%%%%%%%%%%%%%%%%%%%%%%%%%%%%%%%%%%%%%%%%%%%%%%%%%%%%%%%%%%%%%%%%%%%%%%%
  \bibitem{DelCima:2011bx}
  O.~M.~Del Cima,
  `` \textit{The Jackiw-Pi model and its symmetries,}''
\doi{ 10.1088/1751-8113/44/35/352001}{ J.\ Phys.\ A {\bf 44}, 352001 (2011)},
   \arXiv{1104.0164}{hep-th}.
%%%%%%%%%%%%%%%%%%%%%%%%%%%%%%%%%%%%%%%%%%%%%%%%%%%%%%%%%%%%%%%%%%%%%%%%%%%%%%%%%%%%%%%%%%%%%%%%%%%%%%%%%%%
   \bibitem{Gupta:2011cta}
  S.~Gupta, R.~Kumar and R.~P.~Malik,
  `` \textit{Superfield approach to nilpotent symmetries in 3D Jackiw-Pi model of massive non-Abelian theory,}''
  \doi{10.1139/cjp-2013-0457}{Can.\ J.\ Phys.\  {\bf 92}, 1033 (2014)},
   \arXiv{1108.1547}{hep-th}.
%%%%%%%%%%%%%%%%%%%%%%%%%%%%%%%%%%%%%%%%%%%%%%%%%%%%%%%%%%%%%%%%%%%%%%%%%%%%%%%%%%%%%%%%%%%%%%%%%%%%%%%%%%%
\bibitem{DelCima:2012bm}
  O.~M.~Del Cima,
  `` \textit{The Jackiw-Pi model: classical theory,}''
\doi{10.1016/j.physletb.2013.02.016}{Phys.\ Lett.\ B {\bf 720}, 254 (2013)},
   \arXiv{1209.5476}{hep-th}.
%%%%%%%%%%%%%%%%%%%%%%%%%%%%%%%%%%%%%%%%%%%%%%%%%%%%%%%%%%%%%%%%%%%%%%%%%%%%%%%%%%%%%%%%%%%%%%%%%%%%%%%%%%%
\bibitem{Kumar:2015mnp}
  R.~Kumar and A.~Lahiri,
  `` \textit{Dimensional reduction of four-dimensional topologically massive gauge theory,}''
\arXiv{1507.05771}{hep-th}.
%%%%%%%%%%%%%%%%%%%%%%%%%%%%%%%%%%%%%%%%%%%%%%%%%%%%%%%%%%%%%%%%%%%%%%%%%%%%%%%%%%%%%%%%%%%%%%%%%%%%%%%%%%%
  \bibitem{Nishino:2015hha}
  H.~Nishino and S.~Rajpoot,
  `` \textit{Extended Jackiw-Pi model and its supersymmetrization,}''
    \doi{10.1016/j.physletb.2015.05.029}{Phys.\ Lett.\ B {\bf 747}, 93 (2015)},
   \arXiv{1510.06713}{hep-th}.
%%%%%%%%%%%%%%%%%%%%%%%%%%%%%%%%%%%%%%%%%%%%%%%%%%%%%%%%%%%%%%%%%%%%%%%%%%%%%%%%%%%%%%%%%%%%%%%%%%%%%%%%%%%
\bibitem{Nikoofard:2016tzz}
  V.~Nikoofard and E.~M.~C.~Abreu,
  `` \textit{New aspects of quantization of Jackiw-Pi model: field-antifield formalism and noncommutativity,}''
\doi{10.1002/andp.201600119}{Annalen Phys.\  {\bf 528}, 705 (2016)},
\arXiv{1601.00860}{hep-th}.
%%%%%%%%%%%%%%%%%%%%%%%%%%%%%%%%%%%%%%%%%%%%%%%%%%%%%%%%%%%%%%%%%%%%%%%%%%%%%%%%%%%%%%%%%%%%%%%%%%%%%%%%%%%
\bibitem{Deser:2012ci}
  S.~Deser, S.~Ertl and D.~Grumiller,
  `` \textit{Canonical bifurcation in higher derivative, higher spin, theories,}''
  \doi{10.1088/1751-8113/46/21/214018}{J.\ Phys.\ A {\bf 46}, 214018 (2013)},
   \arXiv{1208.0339}{hep-th}.
%%%%%%%%%%%%%%%%%%%%%%%%%%%%%%%%%%%%%%%%%%%%%%%%%%%%%%%%%%%%%%%%%%%%%%%%%%%%%%%%%%%%%%%%%%%%%%%%%%%%%%%%%%%
\bibitem{Jurco:2001rq}
  B.~Jurco, L.~Moller, S.~Schraml, P.~Schupp and J.~Wess,
  `` \textit{Construction of nonAbelian gauge theories on noncommutative spaces,}''
  \doi{10.1007/s100520100731}{Eur.\ Phys.\ J.\ C {\bf 21}, 383 (2001)},
       \arXivOld{hep-th/0104153}.
%%%%%%%%%%%%%%%%%%%%%%%%%%%%%%%%%%%%%%%%%%%%%%%%%%%%%%%%%%%%%%%%%%%%%%%%%%%%%%%%%%%%%%%%%%%%%%%%%%%%%%%%%%%
 %%%%%%%%%%%%%%%%%%%%%%%%%%%%%%%%%%%%%%%%%%%%%%
   \bibitem{ref16}
  M.~R.~Douglas and N. Nekrasov,
  ``\textit{Noncommutative field theory},''
  \doi{ 10.1103/RevModPhys.73.977}{Rev. Mod. Phys. {\bf 73}, 977 (2001)},
      \arXivOld{hep-th/0106048};
%%%%%%%%%%%%%%%%%%%%%%%%%%%%%%%%%%%%
 %%%%%%%%%%%%%%%%%%%%%%%%%%%%%%%%%%%%%%%%%%%%%%
   \bibitem{ref17}
   R.~J.~Szabo,
  ``\textit{Quantum field theory on \NC spaces},''
 \doi{10.1016/S0370-1573(03)00059-0}{Phys. Rep. {\bf 378}, 207 (2003)},
 \arXivOld{hep-th/0109162}.
%%%%%%%%%%%%%%%%%%%%%%%%%%%%%%%%%%%%
 %%%%%%%%%%%%%%%%%%%%%%%%%%%%%%%%%%%%%%%%%%%%%%
   \bibitem{ref18}
I.~Hinchliffe, N.~Kersting and Y.~L.~Ma,
  ``\textit{Review of the phenomenology of \NC geometry},''
  \doi{ 10.1142/S0217751X04017094}{Int.\ J.\ Mod.\ Phys.\ A {\bf 19}, 179 (2004)},
\arXivOld{hep-ph/0205040}.
  %%%%%%%%%%%%%%%%%%%%%%%%%%%%%%%%%%%%%%%%%%%
\bibitem{ref23}
G.~Amelino-Camelia,
  ``\textit{Quantum-Spacetime Phenomenology,}''
    \doi{10.12942/lrr-2013-5}{Living Rev.\ Relativity  {\bf 16}, 5 (2013)},
  \arXiv{0806.0339}{gr-qc}.
%%%%%%%%%%%%%%%%%%%%%%%%%%%%%%%%%%%%%%%%%%%%%%%%%%%%%%%%%%%%%%%%%%%%%%%%%%%%%%%%%%%%%%%%%%%%%%%%%%%%%%%%%%%
\bibitem{fradkin}
  E.~Fradkin, V.~Jejjala and R.~G.~Leig,
  ``\textit{Noncommutative Chern-Simons for the quantum Hall
system and duality,},''
  \doi{10.1016/S0550-3213(02)00616-8}{Nucl.\ Phys.\ B {\bf 642}, 483 (2002)},
      \arXivOld{cond-mat/0205653}.
  %%%%%%%%%%%%%%%%%%%%%%%%%%%%%%%%%%%%%%%%%%%%%%%%%%%%%%%%%%%%%%%%%%%%%%%%%%%%%%%%%%%%%%%%%%%%%%%%%%%%%%%%%%%
  \bibitem{Ghasemkhani:2015tqu}
  M.~Ghasemkhani and R.~Bufalo,
  `` \textit{Noncommutative Maxwell-Chern-Simons theory: One-loop dispersion relation analysis,}''
   \doi{10.1103/PhysRevD.93.085021}{Phys.\ Rev.\ D {\bf 93}, 085021 (2016)},
  \arXiv{1512.04094}{hep-th}.
 %%%%%%%%%%%%%%%%%%%%%%%%%%%%%%%%%%%%%%%%%%%%%%%%%%%%%%%%%%%%%%%%%%%%%%%%%%%%%%%%%%%%%%%%%%%%%%%%%%%%%%%%%%%%
\bibitem{Matusis:2000jf}
  A.~Matusis, L.~Susskind and N.~Toumbas,
  `` \textit{The IR/UV connection in the noncommutative gauge theories,}''
  \doi{10.1088/1126-6708/2000/12/002}{JHEP {\bf 0012}, 002 (2000)}, \arXivOld{hep-th/0002075}.
%%%%%%%%%%%%%%%%%%%%%%%%%%%%%%%%%%%%%%%%%%%%%%%%%%%%%%%%%%%%%%%%%%%%%%%%%%%%%%%%%%%%%%%%%%%%%%%%%%%%%%%%%%%
\bibitem{Martin:2017nhg}
  C.~P.~Martin, J.~Trampetic and J.~You,
  `` \textit{Quantum noncommutative ABJM theory: first steps,}''
    \doi{10.1007/JHEP04(2018)070}{JHEP {\bf 04},  070 (2018)},
  \arXiv{1711.09664}{hep-th}.
  %%%%%%%%%%%%%%%%%%%%%%%%%%%%%%%%%%%%%%%%%%%%%%%%%%%%%%%%%%%%%%%%%%%%%%%%%%%%%%%%%%%%%%%%%%%%%%%%%%%%%%%%%%%

    \end{thebibliography}
\end{document}